\documentclass[12pt,a4paper]{article}

\newcommand{\la}{\lambda}
\newcommand{\ga}{\gamma}
\newcommand{\pa}{\partial}

\begin{document}

\begin{flushright}
hep-th/0503239
\end{flushright}
\vspace{1.8cm}

\begin{center}
 \textbf{\Large Wound and Rotating Strings in $AdS_5 \times S^5$}
\end{center}
\vspace{1.6cm}
\begin{center}
 Shijong Ryang
\end{center}

\begin{center}
\textit{Department of Physics \\ Kyoto Prefectural University of Medicine
\\ Taishogun, Kyoto 603-8334 Japan}
\par
\texttt{ryang@koto.kpu-m.ac.jp}
\end{center}
\vspace{2.8cm}
\begin{abstract}
By using the non-diagonal uniform gauge for the Nambu-Goto string action
we derive a gauge-fixed Hamiltonian of a square-root form  for the closed
string in $AdS_5 \times S^5$ which is wound and rotating in an angular 
direction in $S^5$. From the Nambu-Goto string action 
using a non-diagonal gauge we construct a solution describing
a wound string which rotates in the same angular direction as the winding
direction. The relation between energy and angular momentum of the 
string solution is characterized by the
winding number and the bending number, and becomes linear  
in the large angular momentum limit. The small angular momentum
limit is compared with the strong coupling limit of the gauge-fixed
Hamiltonian. We analyze a wound string solution which is rotating in the
different angular direction from the winding direction.
\end{abstract}
\vspace{3cm}
\begin{flushleft}
March, 2005 
\end{flushleft}

\newpage
\section{Introduction}

The AdS/CFT correspondence \cite{MGW} has more and more revealed the deep
relations between the weak coupling gauge theory and the strong coupling
string theory, and vice versa. The obstacle to verify this conjecture
beyond the supergravity approximation is the difficulty of quantizing the
superstring theory in the $AdS_5 \times S^5$ background. However, the 
solvability of the string theory in the pp-wave background \cite{MT}
has presented an important base to give an interesting proposal that
the energies of specific free string excited states can be matched with
the perturbative scaling dimensions of gauge invariant near-BPS operators
with large R-charge in the BMN limit for the $\mathcal{N}=4$ SU(N) super
Yang-Mills (SYM) theory \cite{BMN}. The BMN result has been interpreted 
as the semiclassical quantization of nearly point-like string with
large angular momentum along the central circle of $S^5$ \cite{GKP}.
Various semiclassical extended string configurations with several
large angular momenta in $AdS_5 \times S^5$ which usually go beyond
the BMN scaling have been constructed extensively to study the 
AdS/CFT correspondence for non-BPS states \cite{FRM,FT,AFR} and 
reviewed in \cite{AT}. 

There has been an important step that 
the dilatation operator for the SO(6)
sector in the planar $\mathcal{N}=4$ SYM theory can be interpreted as a
Hamiltonian of an integrable spin chain in the one-loop approximation 
\cite{MZ}. This observation has been generalized to studies of the 
complete dilatation operator \cite{BS} and the higher loop integrability
\cite{BKS}( see \cite{NB} for review ). The anomalous dimension of 
gauge invariant composite operator has 
been computed by using the Bethe ansatz
for diagonalization of the dilatation operator. The one-loop Bethe ansatz
in the SU(2) sector has been shown to match with the prediction of the
one-loop semiclassical string approach \cite{BMS}. Using the 
higher-loop Inozemtsev-Bethe ansatz, the matching has been shown to hold
at two loops \cite{SS}. At the one-loop and two-loop orders there has
been a general proof of the equivalence between the Bethe equation for
the spin chain in the SU(2) sector and the classical Bethe equation
for the classical $AdS_5 \times S^5$ string sigma model \cite{KMM},
whose approach has been further extended to the other sectors such
as SL(2), SO(6) and so on \cite{KZ}. The matching between the anomalous
dimensions of gauge operators and the energies of dual semiclassical 
strings has been further confirmed up to two loops by constructing
the conserved commuting charges and extending the other sectors 
\cite{AS,EMZ}. The other relevant aspects of the gauge/string duality
have been investigated in \cite{KL}. The gauge/string duality has been 
also presented at the level of equations of motion \cite{AM},
and at the level of effective action \cite{MK} where an interpolating
spin chain sigma model action describing the continuum limit of the spin
chain in the coherent basis is constructed.
The approach based on the spin chain sigma model has been further
investigated \cite{KRT,DR,AAT}.

However, the matching has broken down at three loops \cite{SS,AS}.
A similar three-loop disagreement has also appeared \cite{CLM} in the
string analysis of the $1/J$ quantum corrections to the near BMN states
\cite{PR} where $J$ is a single $S^5$ angular momentum.  
A fully reliable comparison requires the complete summation of the gauge
theoretic perturbative expansion, that is, the spectrum of all-loop
dilatation operator. In the closed SU(2) sector the three-loop dilatation
operator coincides with the Hamiltonian of the integrable Inozemtsev 
long-range spin chain \cite{SS}, which however 
violates the BMN scaling at four loops.
A new long-range spin chain which is different from the Inozemtsev 
long-range spin chain at four loops has 
been presented \cite{BDS}, where the all-loop 
asymptotic Bethe ansatz is proposed to recover the string-theoretic
BMN formula. This all-loop ansatz has been further developed 
\cite{AFS,MS} by adding the factorized scattering terms for the local
excitations to make a particular discretization of the classical
continuous Bethe equation for classical strings in \cite{KMM}.
This quantum Bethe ansatz \cite{AFS} reprodeces the $1/J$ quantum 
corrections \cite{CLM} to the energies of BMN string states and recovers
the $\la^{1/4}$ asymptotics \cite{GKP} of anomalous dimensions in the
strong limit of the 't Hooft coupling constant $\la$, that are
entirely determined by the novel scattering terms. This ansatz has
a corresponding perturbative spin chain Hamiltonian 
at weak coupling \cite{NBT}.

In ref. \cite{AF} from the phase-space approach to the Polyakov action 
for the bosonic closed string wound around an angular direction $\phi$
in $S^5$ and moving in $AdS_5 \times S^5$, the Hamiltonian has been
constructed to be of a square-root form and expressed in terms of 
an angular momentum $J$ associated with the $\phi$ direction and
the string tension $\sqrt{\la}$. The physical Hamiltonian is derived
by choosing the non-diagonal uniform gauge which is related to the
static gauge by a 2d duality transformation \cite{KRT}.
In the sector of strings which do not wind around the $\phi$ direction the
large $J$ expansion of the physical Hamiltonian with the BMN coupling
$\la' = \la/J^2$ fixed reproduces the plane-wave Hamiltonian and the 
$1/J$ and $1/J^2$ corrections in \cite{CLM,IS} and in the strong 
coupling limit the $\la^{1/4}$ behavior is recovered. In the sector of
strings which wind around the $\phi$ direction the string energy scales
as $\sqrt{\la}$ in the strong coupling limit. The gauge-fixed physical
Hamiltonian has been shown to be integrable by constructing the
corresponding Lax representation. The integrability properties of 
superstring in $AdS_5 \times S^5$ have been 
further explored to investigate the Lax connections, 
their monodromies and the corresponding conserved
charges \cite{BPR}. On the other hand there have been various studies 
of comparing the quantum corrections to spinning string solutions in 
$AdS_5 \times S^5$ with the finite size corrections to the Bethe 
equations \cite{LZ}.
 
A non-diagonal gauge has been used for the Nambu-Goto action of the 
bosonic closed string moving in $AdS_5$ to construct a new type of
string solution which is rotating and wound one time around an
angular direction of $AdS_3$ embedded in $AdS_5$, and is stretched along
the radial direction of $AdS_3$ with cusps or spikes \cite{MKI}.
In the large angular momentum limit the energy of string has been shown to
be equal to the angular momentum with the sub-leading logarithmic
anomalous term, where the motion of spikes is described by an effective
classical mechanics that agrees with a coherent state description of the
operators in the spin chain.

Using the non-diagonal uniform gauge we will analyze the Nambu-Goto
string action to reconstruct the physical Hamiltonian for the closed
string moving in $AdS_5 \times S^5$ and wound $M$ times around the 
$\phi$ direction in $S^5$. It will be expressed directly in a square-root
form. In order to understand the role of the winding number, from the 
Nambu-Goto string action we will choose a non-diagonal gauge to construct
a solution describing the string which stays at the origin of $AdS_5$ and
is rotating and wound $M$ times in the $\phi$ direction and stretched
along an angular direction in $S^5$. 
The other stretched string solution wound
in the $\phi$ direction but rotating in the other angular direction 
in $S^5$ will be constructed. In certain limits specified by the large
angular momentum or the small one, the relations between the energy  
and the relevant angular momentum will be extracted for
both string solutions.   

\section{Physical Hamiltonian for wound and rotating string}

We consider a closed string moving in $AdS_5 \times S^5$ with metric 
in the global coordinate
\begin{equation}
ds^2 = - \cosh^2\rho dt^2 + d\rho^2 + \sinh^2\rho d\Omega_3^2 + 
\cos^2\ga d\phi^2 + d\ga^2 + \sin^2\ga d\tilde{\Omega}_3^2,
\label{met}\end{equation}
where the radius of each subspace is unit and $d\Omega_3^2$, 
$d\tilde{\Omega}_3^2$ are metrics of separate three-spheres parametrized
by the angles $\beta_k, \tilde{\beta}_k, k = 1, 2, 3$ respectively.
Through the reparametrization
\begin{equation}
\cosh\rho = \frac{1 + \frac{z^2}{4}}{1 - \frac{z^2}{4}}, \hspace{1cm}
\cos\ga = \frac{1 - \frac{y^2}{4}}{1 + \frac{y^2}{4}}
\end{equation}
this metric becomes
\begin{equation}
ds^2 = - \left( \frac{1 + \frac{z^2}{4}}{1 - \frac{z^2}{4}} \right)^2
dt^2 + \left( \frac{1 - \frac{y^2}{4}}{1 + \frac{y^2}{4}} \right)^2
d\phi^2 + \frac{dz_idz_i}{(1 - \frac{z^2}{4})^2} +
\frac{dy_idy_i}{(1 + \frac{y^2}{4})^2},
\label{dst}\end{equation}
where $z^2 = z_iz_i, y^2 = y_iy_i$ with $i = 1, \cdots, 4$, and 
$(\rho, \beta_k), (\ga, \tilde{\beta}_k)$ are replaced by the four
Cartesian coordinates $z_i, y_i$ respectively. In the metric (\ref{dst})
the translation invariance in $t$ and $\phi$ and the 
$SO(4) \times SO(4)$ symmetry of the transverse coordinates $z_i, y_i$
are manifest. We analyze the closed string wound $M$ times 
in the $\phi$ direction so that the angular coordinate $\phi$ 
satisfies the constraint
\begin{equation}
\phi(2\pi) - \phi(0) = -2\pi M, \; M \in Z.
\label{win}\end{equation}

The Nambu-Goto string action is used to study the string propagation
in the metric (\ref{dst}). We choose the non-diagonal uniform gauge
which is considered in \cite{KRT} and provided by the two conditions
\begin{equation}
t = \tau, \hspace{1cm} p_{\phi} = J,
\label{uni}\end{equation}
where $p_{\phi}$ is the canonical momentum conjugate to $\phi$.
The uniform gauge uses the gauge freedom to require that the
target-space time would coincide with the world-sheet time, and that
the angular momentum $J$ would be homogeneously distributed along
the string. Through the first condition $t = \tau$ the relevant
Nambu-Goto  action is given by
\begin{eqnarray}
S_{NG} &=& -\frac{\sqrt{\la}}{2\pi} \int d\tau d\sigma
\sqrt{D},  \label{sng} \\
D &=& F \left( G_{tt} - \frac{\dot{z}_i^2}{(1 - \frac{z^2}{4})^2} - 
G_{\phi\phi}\dot{\phi}^2 - \frac{\dot{y}_i^2}{(1 + \frac{y^2}{4})^2}
\right) + H^2, \nonumber
\end{eqnarray}
where the string tension $\sqrt{\la}$ is expressed in terms of
radius $R$ of $AdS_5$ and $S^5$ as $\sqrt{\la} = R^2/\alpha'$ and
\begin{eqnarray}
F &=& \frac{{z'}_i^2}{(1 - \frac{z^2}{4})^2} + G_{\phi\phi}{\phi'}^2 
+ \frac{{y'}_i^2}{(1 + \frac{y^2}{4})^2}, \nonumber \\
H &=& \frac{\dot{z}_iz'_i}{(1 - \frac{z^2}{4})^2} + G_{\phi\phi}\dot{\phi}
\phi' + \frac{\dot{y}_iy'_i}{(1 + \frac{y^2}{4})^2}
\end{eqnarray}
with two functions
\begin{equation}
G_{tt} = \left( \frac{1 + \frac{z^2}{4}}{1 - \frac{z^2}{4}} \right)^2,
\hspace{1cm} G_{\phi\phi} = \left( 
\frac{1 - \frac{y^2}{4}}{1 + \frac{y^2}{4}} \right)^2,
\end{equation}
where the ``dot" and ``prime" denote differentials with respect to
$\tau$ and $\sigma$ respectively. 

The canonical momenta $p_{z_i}, p_{y_i}$ conjugate to 
$z_i, y_i$ and $p_{\phi}$ are given by
\begin{eqnarray}
p_{z_i} &=& \sqrt{\frac{\la}{D}} \frac{F\dot{z}_i - Hz'_i}
{(1 - \frac{z^2}{4})^2}, \hspace{1cm} 
p_{y_i} = \sqrt{\frac{\la}{D}} \frac{F\dot{y}_i - Hy'_i}
{(1 + \frac{y^2}{4})^2}, \nonumber \\
p_{\phi} &=& \sqrt{\frac{\la}{D}}G_{\phi\phi}(F\dot{\phi} - H\phi').
\label{mom}\end{eqnarray}
Instead of eliminating the velocities we substitute the momenta 
(\ref{mom}) into the Hamiltonian density 
$\mathcal{H} = \dot{z}_ip_{z_i} + \dot{\phi}p_{\phi} +
\dot{y}_ip_{y_i} - \mathcal{L}$ 
with $\mathcal{L} = - \sqrt{\la D}$ to have
\begin{equation}
\mathcal{H} =\sqrt{\frac{\la}{D}} FG_{tt}.
\label{hfg}\end{equation}
In the expression $D$ of (\ref{sng}) here we can eliminate the velocities
by taking account of the relations in (\ref{mom}) to obtain
\begin{equation}
D = \frac{\la G_{tt}F^2}{\la F + (1 - \frac{z^2}{4})^2p_{z_i}^2 +
\frac{p_{\phi}^2}{G_{\phi\phi}} + (1 + \frac{y^2}{4})^2p_{y_i}^2 
- \frac{1}{F}(z'_ip_{z_i} + \phi'p_{\phi} + y'_ip_{y_i})^2 } .
\label{dgf}\end{equation}
Combining (\ref{hfg}) and (\ref{dgf}) we derive a Hamiltonian density
of a square-root form
\begin{equation}
\mathcal{H} = \sqrt{G_{tt} \left[
\la F + \left(1 - \frac{z^2}{4}\right)^2p_{z_i}^2 +
\frac{p_{\phi}^2}{G_{\phi\phi}} + 
\left(1 + \frac{y^2}{4}\right)^2p_{y_i}^2 
- \frac{1}{F}(z'_ip_{z_i} + \phi'p_{\phi} + y'_ip_{y_i})^2 \right] }.
\end{equation}
Specially the last term in the square root proportional to $1/F$ exhibits
the involved rational expression. However, there is a constraint
\begin{equation}
z'_ip_{z_i} + \phi'p_{\phi} + y'_ip_{y_i} = 0,
\label{con}\end{equation}
which holds identically through the momentum expressions of (\ref{mom}).
By using the constraint (\ref{con}) to eliminate $\phi'$ and taking 
account of the second condition in (\ref{uni}) $p_{\phi} = J$ here
we obtain a positive Hamiltonian density expressed in terms of
the physical variables as
\begin{eqnarray}
\mathcal{H} &=& \Biggl[ \frac{G_{tt}}{G_{\phi\phi}}J^2 + 
\frac{\la}{J^2}G_{tt}G_{\phi\phi}
( z'_ip_{z_i} + y'_ip_{y_i})^2 \nonumber \\
&+& G_{tt}\left(1 - \frac{z^2}{4}\right)^2p_{z_i}^2 + \la \frac{G_{tt}}
{(1 - \frac{z^2}{4})^2}{z'}_i^2 + G_{tt}\left(1 + 
\frac{y^2}{4}\right)^2p_{y_i}^2 
+ \la \frac{G_{tt}}{(1 + \frac{y^2}{4})^2}{y'}_i^2 \Biggr]^{1/2}.
\label{sqh}\end{eqnarray}
This expression parametrized by the string tension $\sqrt{\la}$ and a 
single $S^5$ angular momentum $J$ agrees with the result in \cite{AF},
where the physical Hamiltonian of a square-root form was obtained from
the Polyakov string action by using the non-diagonal uniform gauge
and it was shown that in the strong coupling limit $\la \rightarrow
\infty$ with $J$ fixed, the string energy for $M =0$ scales as
$\la^{1/4}$, while that for $M \neq 0$ scales as $M\sqrt{\la}$. 

\section{Energy-spin relations of wound and rotating string solutions}

Based on the Nambu-Goto string action we study a closed string located
at the center $\rho =0$ of $AdS_5$ and rotating in $S^5$.
The metric of $AdS_5 \times S^5$ is given by (\ref{met}) with 
$d\tilde{\Omega}_3^2 = d\psi^2 + \cos^2\psi d\varphi_1^2 +
\sin^2\psi d\varphi_2^2$. We choose world-sheet coordinates 
in such a way that 
\begin{equation}
t = \tau, \hspace{1cm} \phi = \omega \tau + M\sigma,
\label{nod}\end{equation}
where $M$ is the winding number corresponding to the integer $M$ in
(\ref{win}). In order to express a configuration that a closed string is
rotating in the $\varphi_1$ direction as well as the $\phi$ direction
and stretched along the angular coordinate $\ga$, we make 
further the following ansatz
\begin{equation}
\varphi_1 = \omega_1\tau, \; \ga(\sigma) = \ga(\sigma + 2\pi), \;
\psi =  \varphi_2 = 0.
\end{equation}
In this wound and rotating string configuration with $M \neq 0$ we note
$\partial_{\tau}X^{\mu}\partial_{\sigma}X^{\nu}G_{\mu\nu} \neq 0$ and 
therefore the gauge choice (\ref{nod}) is not the diagonal conformal
gauge. The non-diagonal gauge choice (\ref{nod}) with an $AdS_5$ angular
coordinate $\phi$ was taken to construct a wound and rotating string
with spikes in $AdS_5$ in ref. \cite{MKI}. 
From the action (\ref{sng}) with
\begin{equation}
D = - \ga'^2( -1 + \sin^2\ga \dot{\varphi}_1^2 + \cos^2\ga\dot{\phi}^2 )
+ {\phi'}^2\cos^2\ga( 1 - \sin^2\ga \dot{\varphi}_1^2 )
\label{drm}\end{equation}
the equation of motion for $\phi$ yields
\begin{equation}
\pa_{\sigma}\left( \frac{( 1 - \omega_1^2\sin^2\ga )\cos^2\ga}
{\sqrt{D}} \right) = 0,
\end{equation}
while that for $\varphi_1$ is trivially satisfied. 
Therefore we have a relation expressed in terms of an integration
constant $k$ as
\begin{equation}
\sqrt{D} = k( 1 - \omega_1^2\sin^2\ga )\cos^2\ga,
\label{sqd}\end{equation}
which combines with (\ref{drm}) to give 
\begin{equation}
\ga' = \cos\ga \frac{ \sqrt{( 1 - \omega_1^2\sin^2\ga )
[k^2\cos^2\ga(1 - \omega_1^2\sin^2\ga ) - M^2]} }
{ \sqrt{1 - \omega_1^2\sin^2\ga - \omega^2\cos^2\ga} }.
\label{gam}\end{equation}
The equation of motion for $\ga$ is written by 
\begin{eqnarray}
&\pa_{\sigma}\left( \frac{\ga'}{\sqrt{D}}(1 - \omega_1^2\sin^2\ga
 - \omega^2\cos^2\ga) \right) \nonumber \\
& = \frac{1}{\sqrt{D}}\sin\ga\cos\ga \left( (\omega^2 - \omega_1^2)
{\ga'}^2 - M^2(1 + \omega_1^2\cos^2\ga  - \omega_1^2\sin^2\ga) \right).
\label{ega}\end{eqnarray} 
Substituting (\ref{sqd}) and (\ref{gam}) into (\ref{ega}) we can show 
that the involved equation (\ref{ega}) is indeed satisfied.
Thus we obtain the first integral of the equation of motion.

Let us concentrate on a simpler case specified by $\omega \neq 0, 
\omega_1 =0$. We impose a condition $0 < M/k \le 1/\omega \le 1$.
From $\ga' = \cos\ga\sqrt{k^2\cos^2\ga - M^2}
/\sqrt{1-\omega^2 \cos^2\ga}$ 
there appears a range which is bounded by two turning points
as $M/k \le \cos\ga \le 1/\omega$.
Therefore $\ga(\sigma)$ which is the length of a circular arc from
the equator on the unit-radius $S^2$ parametrized by $\ga$ and 
$\phi$, varies as $\sigma$ increases from a minimum value $\ga_- =
\arccos(1/\omega)$ to a maximum vale $\ga_+ = \arccos(M/k)$.
At $\ga = \ga_-$, $\ga'$ diverges indicating the presence of a
sharp wedge in the trajectory of $\ga(\sigma)$ and at $\ga = \ga_+$, 
$\ga'$ vanishes so that there is a smooth peak between wedges.
By gluing $2N$ of the arc segments we construct a string configuration
wound $M$ times around the $\phi$ direction with $N$ wedges.  
The angle difference between the wedge and the peak is given by
\begin{equation}
\Delta \phi \equiv \frac{2\pi M}{2N} = \frac{\cos\ga_+}{\cos\ga_-}
\int_{\ga_-}^{\ga_+}d\ga \frac{1}{\cos\ga} 
\sqrt{\frac{\cos^2\ga_- - \cos^2\ga}{\cos^2\ga - \cos^2\ga_+} },
\end{equation}
while the angular momentum $J$ associated with the rotation in the $\phi$
direction and the energy $E$ of the string are obtained by
\begin{eqnarray}
J &=& \frac{\sqrt{\la}}{2\pi} 2N \int_{\ga_-}^{\ga_+}d\ga \cos\ga
\sqrt{\frac{\cos^2\ga - \cos^2\ga_+}{\cos^2\ga_- - \cos^2\ga} }, \\
E - \omega J &=& \frac{\sqrt{\la}N\omega}{\pi}
\int_{\ga_-}^{\ga_+}d\ga \cos\ga 
\sqrt{\frac{\cos^2\ga_- - \cos^2\ga}{\cos^2\ga - \cos^2\ga_+} }.
\end{eqnarray}
These integrals are expressed in terms of complete elliptic integrals as
\begin{eqnarray}
\Delta\phi &=& \frac{x_-^2}{x_+} \sqrt{ \frac{1-x_+^2}{1- x_-^2} }
\left[ \Pi\left( \frac{q^2}{1- x_-^2}, q \right) - K(q) \right],
\label{dph} \\
J &=& \frac{\sqrt{\la}N}{\pi} x_+ [ K(q) - E(q) ] = 
 \frac{\sqrt{\la}N}{4} x_+ q^2 F\left( \frac{3}{2},\frac{1}{2},2,q^2 
\right), \label{jke} \\
E - \omega J &=& \frac{\sqrt{\la}N}{\pi}\frac{1}{\sqrt{1- x_-^2}}
\left[ x_+ E(q) - \frac{x_-^2}{x_+}K(q) \right],
\label{eoj}\end{eqnarray}
where $x_{\pm} = \sin \ga_{\pm}, \; q = \sqrt{x_+^2 - x_-^2}/x_+$
and $F(\frac{3}{2},\frac{1}{2},2,q^2)$ is the hypergeometric function.
Thus some combinations of the integration constants such as $k$ and 
$\omega$ appear as parameters of the elliptic functions.
From (\ref{jke}) and (\ref{eoj}) the energy expression is extracted as
\begin{equation}
E = \frac{\sqrt{\la}N}{\pi} \frac{x_+}{\sqrt{1- x_-^2}} q^2 K(q).
\label{enk}\end{equation}
Eliminating $x_+$ and $x_-$ in (\ref{dph}), (\ref{jke}) and (\ref{enk}) we
determine the energy $E$ as a function of the angular momentum $J$, 
the winding number $M$ and the bending number $N$. 

We will derive explicit analytic expressions in three different regions.
From (\ref{jke}) and (\ref{enk}) the region of large $E$ and large $J$ as
compared to $\sqrt{\la}$ corresponds to a region $q \approx 1$ which means
that $x_- \approx 0$, that is, $\omega$ approaches 1 from above.
Thus the large $J$ region is specified by $r_- \approx 0$.
Both $E$ and $J$ diverge logarithmically owing to $K(q)$ and are
expressed to the leading order as
\begin{equation}
E \approx J \approx \frac{\sqrt{\la}N}{\pi} x_+ \log \frac{x_+}{x_-},
\end{equation}
while the difference $E - J$ in (\ref{eoj}) becomes finite 
\begin{equation}
E - J \approx E - \omega J \approx \frac{\sqrt{\la}N}{\pi}x_+
\label{ejx}\end{equation}
owing to the factor $x_-^2$ which suppresses the logarithmic divergence
of $K(q)$. Here we write down a formula
\begin{equation}
\Pi(n,q) = K(q) + \frac{2\sqrt{1-{k'}^2\sin^2\psi}}
{{k'}^2\sin 2\psi}\left[ F(\psi,k')K(q) - E(\psi,k')K(q) - F(\psi,k')E(q)
+ \frac{\pi}{2} \right]
\label{pik}\end{equation}
with ${k'}^2 = 1 - q^2, \; n = 1 - {k'}^2\sin^2\psi$.  
For $n \approx q^2 \approx 1$ it approximately reduces to
\begin{equation}
\Pi(n,q) \approx \sqrt{\frac{n}{(1-n)(n-q^2)}}\left( \frac{\pi}{2} - 
\arcsin \sqrt{\frac{1-n}{1-q^2}} \right),
\label{pin}\end{equation}
whose substitution into (\ref{dph}) provides
\begin{equation}
\cos\Delta\phi \approx \sqrt{1 - x_+^2},
\end{equation}
which implies $2M \le N$. Thus $x_+$ is characterized by $M$ and $N$ as
\begin{equation}
x_+ \approx \sin \frac{M\pi}{N}.
\label{xpu}\end{equation}
From (\ref{ejx}) and (\ref{xpu}) we get
\begin{equation}
E - J \approx  
\frac{\sqrt{\la}N}{\pi} \sin\frac{M\pi}{N}
\label{ejs}\end{equation}
in the large angular momentum limit. For the special $2M = N$
case that corresponds to $\ga_+ = \pi/2$, it reduces to
\begin{equation}
E - J \approx \frac{\sqrt{\la}N}{\pi},
\label{ejn}\end{equation}
which, for $N = 2$ with $M = 1$ shows the same expression as that
in \cite{GKP}.  In the general situations there is a difference that
our closed string configuration is a zigzag bending curve surrounding
the north pole on the relevant $S^2(\ga,\phi)$, 
while the closed string presented in \cite{GKP}
is folded on itself along the meridian in a line form.
For instance in the $M = 1$ and $N = 3$ case the string shape
at fixed time starts at the equator point specified by $\ga = \phi = 0$
that is one wedge to reach the maximum peak point $\ga = \pi/3$ with 
$\phi = \pi/3$ and turns around to 
the second wedge $\ga = 0, \; \phi = 2\pi/3$.
This one step is repeated three times to arrive at the starting point.
In the special $M = 1$ and $N = 2$ case the string shape starts
at the same equator point and goes along the meridian to 
reach the north pole and turns around to $\ga = 0, \; \phi = \pi$.
Then it returns back on itself so that this configuration just shows
the string folded in the $\ga$ direction and composed of the four 
segments. Therefore this picture is consistent with the reconstruction
of the energy-spin relation for the four folded string as a special 
case from the general relation (\ref{ejs}). Similarly for the 
$2M = N$ case the string is folded $2N$ times in the $\ga$ direction.

In view of (\ref{jke}) and (\ref{enk}) the small $J$ and small $E$ region
is specified by small $q$. The two parameters $x_+, x_-$ are traded 
for $m, n$ as follows
\begin{equation}
m = q^2 = \frac{x_+^2 - x_-^2}{x_+^2}, \hspace{1cm}
n = \frac{q^2}{1 - x_-^2} = \frac{x_+^2 - x_-^2}{x_+^2(1 - x_-^2)},
\end{equation}
which satisfy $0 < m \le n < 1$ and inversely give 
$x_-^2 = (n - m)/n, \; x_+^2 = (n - m)/(n(1- m))$. The energy,  
the angular momentum and the angle difference
are expressed in terms of $m, n$ as
\begin{eqnarray}
E &=& \frac{\sqrt{\la}N}{2}\sqrt{\frac{m(n - m)}{1- m}}
\left( 1 + \frac{m}{4} + \cdots \right),
\label{emn} \\
J &=& \frac{\sqrt{\la}N}{4}m \sqrt{\frac{n - m}{n(1- m)}}
\left( 1 + \frac{3}{8}m + \cdots \right), \label{jmn}\\
\Delta\phi &=& F(\psi,k')K(q) - E(\psi,k')K(q) - F(\psi,k')E(q)
+ \frac{\pi}{2},
\label{dpp}\end{eqnarray}
where $q = \sqrt{m}, k' = 1 - m$ and
\begin{equation}
\sin^2 \psi = \frac{1 - n}{1 - m}.
\label{sip}\end{equation}
There are two different small $q$ regions: one region is specified
by $m/n \ll 1$ and $m \ll 1$, and the other one by 
$n \approx m \ll 1 - m/n \ll 1$. 

Let us consider the parameter region specified by $m \ll n \ll 1$
that is more restricted than the former one $m \ll n < 1$.
Since this parameter region means $x_+ \approx x_- \approx 1$, that is,
$\ga_+ \approx \ga_- \approx \pi/2$, the rotating and bending string
is so located near the north pole that it is of a small size.
From (\ref{emn}) and (\ref{jmn}) we make the leading estimations such as
$E \approx \sqrt{\la}N\sqrt{mn}/2$ and $J \approx \sqrt{\la}Nm/4$
so that small $E$ and small $J$ are characterized by small $m$.
The elimination of $m$ yields
\begin{equation}
E^2 \approx Nn\sqrt{\la}J,
\label{enl}\end{equation}
which however, includes a parameter $n$, that will be expressed 
in terms of $J, M$ and $N$ through (\ref{dpp}). The relation 
(\ref{sip}) gives $\psi \approx \pi/2$ to the leading 
order, from which the $\Delta\phi$ in (\ref{dpp}) becomes
 $K(k')K(q)- E(k')K(q)- K(k')E(q) + \pi/2$, 
that is identically zero. The expansion of (\ref{sip})
around $\psi =\pi/2$ gives here for $n \ll 1, \; m \ll 1$
\begin{equation}
(\Delta\psi)^2 \approx n - m + \frac{1}{3}( n^2 + mn - 2m^2 ), 
\hspace{1cm} \psi \equiv \frac{\pi}{2} - \Delta\psi 
\label{dpm}\end{equation}
and the expansion of (\ref{dpp}) yields
\begin{equation}
\Delta\phi \approx \frac{\pi}{4} q\Delta\psi,
\label{phq}\end{equation}
where we have used 
\begin{eqnarray}
\pa_{\psi}F(\psi,k') &=& (1 - {k'}^2\sin^2\psi )^{-1/2}, \hspace{1cm}
\pa_{\psi}^2F(\psi,k')|_{\psi=\pi/2} = 0, \nonumber \\
\pa_{\psi}E(\psi,k') &=& (1 - {k'}^2\sin^2\psi )^{1/2}, \hspace{1cm}
\pa_{\psi}^2E(\psi,k')|_{\psi=\pi/2} = 0.
\end{eqnarray}
The leading value $n \approx (4M/N)^2/m \approx 4M^2\sqrt{\la}/JN$
determined from (\ref{phq}) makes (\ref{enl}) change into
$E^2 \approx 4M^2\la$. By estimating the sub-leading term from
(\ref{phq}), (\ref{emn}) and (\ref{jmn}) we obtain
\begin{equation}
E^2 \approx 4M^2 \left(\la + \frac{14}{3} \frac{J}{N} \sqrt{\la} \right).
\end{equation}
We note that the first term corresponds to the $M\sqrt{\la}$ term which is
determined from (\ref{sqh}) in the strong coupling limit 
for $M \ne 0$ \cite{AF}.

Here we consider the other parameter region $n \approx m \ll 1 - m/n
\ll 1$ which implies $x_+ \approx x_- \ll 1$, that is,  
$\ga_+ \approx \ga_- \approx 0$. Therefore the rotating string is located
near the equator and wound nearly around a central circle of $S^5$
so that it is not of a small size although $J$ is small.
For this particular parameter region, the eqs. (\ref{emn}) and 
(\ref{jmn}) yield a linear relation, to the leading order
\begin{equation}
E \approx 2J.
\label{etw}\end{equation}
The leading expression of (\ref{dpm}) combines with (\ref{phq})
and (\ref{jmn}) to make (\ref{etw}) change into $E \approx M\sqrt{\la}$.

Now we turn to the other simpler case specified by $\omega_1 \neq 0,
\; \omega = 0$. In this case the gauge choice (\ref{nod}) belongs to
the diagonal conformal gauge. The first integral (\ref{gam}) 
is expressed as 
\begin{eqnarray}
\ga' &=& \cos\ga \sqrt{k^2\cos^2\ga(1- \omega_1^2\sin^2\ga) - M^2}
= k\omega_1\cos\ga \sqrt{(\cos^2\ga - \alpha_+)(\cos^2\ga + \alpha_-)},
\nonumber \\
\alpha_{\pm} &=& \frac{ \sqrt{(\omega_1^2 -1)^2 + 4\omega_1^2M^2/k^2}
\pm (\omega_1^2 -1)}{2\omega_1^2}.
\label{alp}\end{eqnarray}
We impose a condition $0<\alpha_- \le \alpha_+ \le 1$ which leads to
$M \le k$ with $\omega_1 \ge 1$. Therefore $\ga(\sigma)$ on 
$S^2(\ga,\phi)$ embedded in $S^5$, varies from zero where $\pa_{\phi}\ga$
takes $\sqrt{(k/M)^2 -1}$ to a maximum value  $\ga_+ = \arccos
\sqrt{\alpha_+}$ where $\pa_{\phi}\ga$ vanishes, and $\ga(\sigma)$ returns
back to zero where $\pa_{\phi}\ga = -\sqrt{(k/M)^2 -1}$. 
The $2N$ gluing of the arc segments yields 
a string configuration wound $M$ times around the
$\phi$ direction with $N$ wedges. The acute angle of a wedge is locally
chraracterized by $2(\pi/2 - \arctan\sqrt{(k/M)^2 -1})$.
From the view point of the string configuration on the other embedded
two-sphere $S^2(\ga,\varphi_1)$ parametrized by $\ga$ 
and $\varphi_1$, $\ga(\sigma)$ which is
the length of circular arc from the north pole on it varies from zero
to a maximum value $\ga_+$ and returns back on itself.
The angle difference between the wedge and the peak on $S^2(\ga,\phi)$
is expressed as
\begin{equation}
\Delta\phi \equiv \frac{2\pi M}{2N} = \frac{M}{k\omega_1}\int_0^{\ga_+}
d\ga \frac{1}{\cos\ga \sqrt{(\cos^2\ga - \alpha_+)
(\cos^2\ga + \alpha_-)}},
\label{omf}\end{equation}
while the angular momentum coming fron the rotation in the $\varphi_1$
direction and the energy of the string are described by
\begin{eqnarray}
J_1 &=& \frac{\sqrt{\la}}{2\pi} 2N \int_0^{\ga_+}
d\ga \frac{\sin^2\ga\cos\ga}
{\sqrt{(\cos^2\ga - \alpha_+)(\cos^2\ga + \alpha_-)}}, 
\label{jfn} \\
E - \omega_1J_1 &=& \frac{\sqrt{\la}N}{\pi\omega_1}
\int_0^{\ga_+}d\ga \frac{\cos\ga(1- \omega_1^2\sin^2\ga)}
{\sqrt{(\cos^2\ga - \alpha_+)(\cos^2\ga + \alpha_-)}}.
\label{ejf}\end{eqnarray}
From (\ref{alp}) the two parameters $\omega_1, k$ are traded for 
$\alpha_+, \alpha_-$ so that $\omega_1, k$ in (\ref{omf}) and (\ref{ejf})
are implicitly expressed in terms of $\alpha_+, \alpha_-$.
The integrations in (\ref{omf}) and (\ref{jfn}) are performed as 
\begin{eqnarray}
\Delta\phi &=& \frac{M}{k\omega_1} \frac{1}{\sqrt{1 + \alpha_-}}
\Pi(1 -\alpha_+,q), \label{dpf} \\
J_1 &=& \frac{\sqrt{\la}N}{\pi} \sqrt{1 + \alpha_-} (K(q) - E(q))
= \frac{\sqrt{\la}N}{4} \sqrt{1 + \alpha_-}q^2
F\left(\frac{3}{2},\frac{1}{2},2,q^2 \right),
\label{kef}\end{eqnarray}
where $q^2 = (1 - \alpha_+)/(1 + \alpha_-)$.
A comparison of (\ref{jfn}) with (\ref{ejf}) yields
\begin{equation}
E = \frac{\sqrt{\la}N}{\pi\omega_1} \frac{1}{\sqrt{1 + \alpha_-}}K(q).
\label{enf}\end{equation}
The eqs. (\ref{kef}) and (\ref{enf}) show the similar expressions
to $J$ in (\ref{jke}) and $E$ in (\ref{enk}) respectively, although
the string configurations are different. The difference 
$E - \omega_1J_1$ in (\ref{ejf}) is then provided by
\begin{equation}
E -  \omega_1J_1 = \frac{\sqrt{\la}N}{\pi\omega_1} 
\frac{1}{\sqrt{1 + \alpha_-}} [\omega_1^2( 1 + \alpha_-)E(q) +
(1 - \omega_1^2( 1 + \alpha_-)) K(q) ],
\label{eof}\end{equation}
which resembles (\ref{eoj}).

From (\ref{kef}) and (\ref{enf}) large $J_1$ and large $E$ are achieved by
taking $q$ to be near the critical value 1.
This region is specified by turning $\alpha_+$ and $\alpha_-$ close to
zero, which corresponds to $M/k \approx 0,\; \omega_1 \approx 1$.
Because of the suppression factor $(1 - \omega_1^2( 1 + \alpha_-))$
multiplying $K(q)$ in (\ref{eof}) the difference $E - J_1$ becomes 
finite  
\begin{equation}
E - J_1 \approx E - \omega_1J_1 \approx \frac{\sqrt{\la}N}{\pi},
\label{emj}\end{equation}
which shows the same expression as (\ref{ejn}). 
In the small $\alpha_{\pm}$ region, $0 \approx \alpha_- \le \alpha_+$
where $\alpha_+ \approx 0$ implies $\ga_+ \approx \pi/2$ 
the use of (\ref{pin}) for (\ref{dpf}) gives
\begin{equation}
\cos\Delta\phi \approx \sqrt{ \frac{\alpha_+( 1 + \alpha_- )}
{\alpha_+ + \alpha_-} },
\end{equation}
which also implies $2M \le N$. When $\omega_1$ approaches 1 from above
we have $\cos\Delta\phi \approx 1$, that is, $M/N \ll 1$.
In the special $M = 0$ case the string is not wound in the $\phi$
direction but folded $2N$ times in the $\ga$ direction.
When $N = 2$ with $M = 0$ the energy-spin relation (\ref{emj}) 
again coincides with that for the folded string composed of 
the four segments in ref. \cite{GKP}.
Alternatively if we substitute $M = 0$ directly into (\ref{kef})
and (\ref{enf}) we have
\begin{equation}
J_1 = \frac{\sqrt{\la}N}{\pi}(K(q) - E(q)), \hspace{1cm}
E = \frac{\sqrt{\la}N}{\pi\omega_1} K(q)
\label{moj}\end{equation}
with $q = 1/\omega_1$, which, for $N = 2$, reduces to those 
in ref. \cite{GKP}. 

We consider the opposite parameter region where $J_1$ is small.
From (\ref{kef}) it is specified by small $q$, that is, 
$\alpha_+ \approx 1$, which leads to the lower bound $k \approx M$ 
through (\ref{alp}). Because of $\ga_+ \approx 0$ the rotating string
is wound nearly around the equator. 
We expand $\alpha_{\pm}$ in (\ref{alp}) as $\alpha_+ = 1 - (1+\alpha_-)
q^2 \approx 1 - \delta/(\omega_1^2 + 1), \; \alpha_- \approx (1 - 
\omega_1^2 \delta/(\omega_1^2 + 1))/\omega_1^2$ with
$\delta \equiv 1 - M^2/k^2$. Plugging  the 
following expansion into (\ref{enf})
\begin{equation}
q^2 \approx \frac{4J_1}{\sqrt{\la}N\sqrt{1 + \alpha_-}} - 
\frac{6J_1^2}{\la N^2(1 + \alpha_-)},
\end{equation}
which is derived from (\ref{kef}), we have
\begin{equation}
E \approx  \frac{\sqrt{\la}N}{2\omega_1 \sqrt{1 + \alpha_-}} +
\frac{J_1}{2\omega_1 (1 + \alpha_-)} 
\end{equation}
with $\alpha_- \approx (1 - \omega_1^2(1 + \alpha_-)q^2)/\omega_1^2$.
The remaining condition (\ref{dpf}) is expressed as 
\begin{equation}
\frac{M}{N} \approx \frac{1}{2\omega_1}
\sqrt{ \frac{1 - (\omega_1^2 + 1)(1 + \alpha_-)q^2}{1+\alpha_-} }
\left[ 1 + \left( \frac{1}{4} + \frac{ \sqrt{1+\alpha_-} }{2} \right)
 q^2 \right],
\end{equation}
where the expansion of (\ref{pik}) around $\psi = \pi/2$ is used.
Focusing on $NJ_1/M^2 \ll \sqrt{\la}$ and
the large $N/M$ region that corresponds to the
large $\omega_1$ region we can estimate the energy as
\begin{equation}
E \approx \sqrt{\la} M \left( 1 + \frac{N}{2M^2}
\frac{J_1}{\sqrt{\la}} \right),
\end{equation}
which scales as $\sqrt{\la}M$ in the strong coupling limit. 
For comparison manipulating the equations in (\ref{moj}) we express
the squared energy for the folded string with small angular momentum as
\begin{equation}
E^2 \approx \sqrt{\la} N \left( J_1 + \frac{1}{2}\frac{J_1^2}
{\sqrt{\la}N} + \frac{1}{4}\frac{J_1^3}{(\sqrt{\la}N)^2} \right),
\end{equation}
which shows the Regge behavior and the $\la^{1/4}$ scaling for energy
in the strong coupling limit.

\section{Conclusion}

By choosing the non-conformal gauges for the Nambu-Goto string action we
have analyzed the rotating closed strings moving in $AdS_5 \times S^5$
and wound around a circle of $S^5$ parametrized by $\phi$.
Starting from the Nambu-Goto string action in the suitable 
parametrization of $AdS_5 \times S^5$ we have used the non-diagonal
uniform gauge to construct a Hamiltonian expressed in terms of the
physical variables for the closed string rotating and wound $M$ times
in the same $\phi$ direction. The obtained physical Hamiltonian is 
characterized by the winding number $M$ and 
the angular momentum $J$ associated
with $\phi$, in agreement with the result of ref. \cite{AF} where the
squared Hamiltonian density expressed as $\mathcal{H}^2 = p_t^2$ 
is derived from the Polyakov string action, 
in which expression $p_t$ is the canonical momentum conjuate to 
the global AdS time $t$. We have directly derived the positive 
Hamiltonian of a square-root form, whereas the Polyakov string
approach made a suitable prescription that the negative root of the
equation was consistently picked up as $\mathcal{H} = -p_t$
for positivity of the physical Hamiltonian.

Solving the non-linear equations of motion derived from the
Nambu-Goto string action we have presented a solution
describing a closed string rotating and wound $M$ times in the same
$\phi$ direction, as well as a solution describing a closed
string wound in the $\phi$ direction but rotating in the other
angular direction of $S^5$. The gauge choice for the former 
string configuration is not the diagonal conformal gauge, unlike
the latter one. We have demonstrated that the string energies in certain
limits are explicitly described in terms of the angular momentum $J$,
the winding number $M$ and the bending number $N$.

For the former configuration we have observed that  
in the large angular momentum limit one of the two turning points 
for the shape trajectory of the wound string is
located at the equator of the relevant $S^2$, while the other turning
point is specified by the ratio $M/N$.
For the special $2M = N$ case the
rotating string of the bending hoop configuration changes into that 
of the $2N$ folded line configuration. In the large angular momentum 
limit the energy of the wound and rotating string has been shown to be 
equal to the large angular momentum with the sub-leading 
term which is specified by the winding number $M$ and the bending
number $N$. On the other hand in the small angular momentum region
we have seen that there exists two types of string configurations,
a short string located near the north pole of the $S^2$ and a string 
wound nearly along its equator. We have demonstrated that the energy of
these strings is described by the string tension times the 
winding number $M$ in the small angular momentum limit, whose behavior
is similar to the strong coupling limit of the physical Hamiltonian
for $M \ne 0$ in ref. \cite{AF}.  

For the latter configuration the wound and rotating string in the
large angular momentum limit has been also shown to become
the $2N$ folded and rotating string for $M = 0$, whose energy 
expression coincides with that of string for the
former configuration specially labelled 
with $2M = N$. Taking account of the situation that 
the rotating angular directions of the two configurations
are different from each other, we note that the different behaviors of
$M$ are consistent in such a way that both strings reduce to the
$2N$ folded string in the large angular momentum limit.
In the small angular momentum limit the energy of the latter string
configuration has been also shown to 
be the string tension times the winding
number $M$. It is desirable to investigate what gauge invariant
composite operators correspond to the wound and rotating string
solutions labelled with the winding and bending numbers.


\begin{thebibliography}{99}
\bibitem{MGW} J.M. Maldacena, ``The large N limit of superconformal
field theories and supergravity,'' Adv. Theor. Math. Phys. \textbf{2}
(1998) 231 [hep-th/9711200]; S.S. Gubser, I.R. Klebanov and A.M. Polyakov,
``Gauge theory correlators from non-critical string theory,"
Phys. Lett. \textbf{B428} (1998) 105 [hep-th/9802109]; E. Witten, 
``Anti-de Sitter space and holography,"
Adv. Theor. Math. Phys. \textbf{2} (1998) 253 [hep-th/9802150].
\bibitem{MT} R.R. Metsaev, ``Type IIB Green-Schwarz superstring in
plane wave Ramond-Ramond background," Nucl. Phys. \textbf{B625}
(2002) 70 [hep-th/0112044];  
R.R. Metsaev and A.A. Tseytlin, ``Exactly solvable 
model of superstring in plane wave Ramond-Ramond background," Phys.
Rev. \textbf{D65} (2002) 126004 [hep-th/0202109].
\bibitem{BMN} D. Berenstein, J.M. Maldacena and H. Nastase, 
``Strings in flat space and pp waves from $\mathcal{N}$=4 super
Yang Mills," JHEP \textbf{04} (2002) 013 [hep-th/0202021].
\bibitem{GKP} S.S. Gubser, I.R. Klebanov and A.M. Polyakov,
``A semi-classical limit of the gauge/string correspondence,"
Nucl. Phys. \textbf{B636} (2002) 99 [hep-th/0204051].
\bibitem{FRM} S. Frolov and A.A. Tseytlin, ``Semiclassical 
quantization of rotating superstring in $AdS_5\times S^5$," JHEP
\textbf{06} (2002) 007 [hep-th/0204226];
J.G. Russo, ``Anomalous dimensions in gauge theories
from rotating strings in $AdS_5\times S^5$," JHEP \textbf{06} (2002)
038 [hep-th/0205244];
J.A. Minahan, ``Circular semiclassical string solutions
on $AdS_5\times S^5$," Nucl. Phys. \textbf{B648}
(2003) 203 [hep-th/0209047].
\bibitem{FT} S. Frolov and A.A. Tseytlin, ``Multi-spin string solutions
in $AdS_5 \times S^5$," Nucl. Phys. \textbf{B668} (2003) 77 
[hep-th/0304255]; S. Frolov and A.A. Tseytlin, ``Quantizing three-spin
string solution in $AdS_5 \times S^5$," JHEP \textbf{07} (2003) 016
[hep-th/0306130]; S. Frolov and A.A. Tseytlin, ``Rotating string 
solutions: AdS/CFT duality in non-supersymmetric sectors," Phys. Lett. 
\textbf{B570} (2003) 96 [hep-th/0306143].
\bibitem{AFR} G. Arutyunov, S. Frolov, J. Russo and A.A. Tseytlin, 
``Spinning strings in $AdS_5 \times S^5$ and integrable systems,"
Nucl. Phys. \textbf{B671} (2003) 3 [hep-th/0307191];
G. Arutyunov, J. Russo and A.A. Tseytlin, ``Spinning strings
in $AdS_5\times S^5$: new integrable system relations," 
Phys. Rev. \textbf{D69} (2004) 0806009 [hep-th/0311004].
\bibitem{AT} A.A. Tseytlin, ``Spinning strings and AdS/CFT duality,"
in: Ian Kogan Memorial Volume, ``From Fields to Strings:
Circumnavigating Theoretical Physics,"  M. Shifman, A. Vainshtein,
and J. Wheater, eds. ( World Scientific, 2004 ), hep-th/0311139.
\bibitem{MZ} J.A. Minahan and K. Zarembo, ``The Bethe-ansatz for 
$\mathcal{N} =4$ super Yang-Mills," JHEP \textbf{03} (2003) 013 
[hep-th/0212208].
\bibitem{BS} N. Beisert, ``The complete one-loop dilatation operator
of $\mathcal{N} =4$ super Yang-Mills theory," Nucl. Phys. \textbf{B676}
(2004) 3 [hep-th/0307015]; N. Beisert and M. Staudacher, ``The 
$\mathcal{N}=4$ SYM integrable super spin chain," Nucl. Phys. 
\textbf{B670} (2003) 439 [hep-th/0307042].
\bibitem{BKS} N. Beisert, C. Kristjansen and M. Staudacher, ``The 
dilatation operator of $\mathcal{N} =4$ super Yang-Mills theory," 
Nucl. Phys. \textbf{B664} (2003) 131 [hep-th/0303060];
N. Beisert, ``Higher loops, integrability and the near
BMN limit," JHEP \textbf{09} (2003) 062 [hep-th/0308074];
``The su$(2|3)$ dynamic spin chain," Nucl. Phys. \textbf{B682}
(2004) 487 [hep-th/0310252].
\bibitem{NB} N. Beisert, ``The dilatation operator of 
$\mathcal{N} = 4$ super Yang-Mills theory and integrability," 
Phys. Rept. \textbf{405} (2005) 1 [hep-th/0407277].
\bibitem{BMS} N. Beisert, J.A. Minahan, M. Staudacher and K. Zarembo, 
``Stringing spins and spinning strings," JHEP \textbf{09}
(2003) 010 [hep-th/0306139];
N. Beisert, S. Frolov, M. Staudacher and A.A. Tseytlin, 
``Precision spectroscopy of AdS/CFT," JHEP \textbf{10} (2003) 037 
[hep-th/0308117].
\bibitem{SS} D. Serban and M. Staudacher, ``Planar $\mathcal{N} =4$ 
gauge theory and the Inozemtsev long range spin chain," JHEP \textbf{06}
(2004) 001 [hep-th/0401057].
\bibitem{KMM} V.A. Kazakov, A. Marshakov, J.A. Minahan and K. Zarembo, 
``Classical/quantum integrability in AdS/CFT," JHEP \textbf{05}
(2004) 024 [hep-th/0402207].
\bibitem{KZ} V.A. Kazakov and K. Zarembo, 
``Classical/quantum integrability in non-compact sector of AdS/CFT," 
JHEP \textbf{10} (2004) 060 [hep-th/0410105];
N. Beisert, V.A. Kazakov and K. Sakai, ``Algebraic curve for the SO(6)
sector of AdS/CFT," hep-th/0410253;
S. Sch\"afer-Nameki, ``The algebraic curve of 1-loop planar 
$\mathcal{N} = 4$ SYM," hep-th/0412254.
\bibitem{AS} G. Arutyunov and M. Staudacher, ``Matching higher conserved
charges for strings and spins," JHEP \textbf{03} (2004) 004 
[hep-th/0310182]; ``Two-loop commuting charges and the string/gauge
duality," hep-th/0403077;
J. Engquist,``Higher conserved charges and integrability for
spinning strings in $AdS_5 \times S^5$," JHEP \textbf{04}
(2004) 002 [hep-th/0402092].
\bibitem{EMZ} J. Engquist, J.A. Minahan and K. Zarembo, 
``Yang-Mills duals for semiclassical strings on $AdS_5 \times S^5$,''
 JHEP \textbf{11} (2003) 063 [hep-th/0310188]; C. Kristjansen,
``Three-spin strings on $AdS_5 \times S^5$ from $\mathcal{N} =4$ SYM," 
Phys. Lett. \textbf{B586} (2004) 106 [hep-th/0402033];
 M. Smedback, ``Pulsating strings on $AdS_5 \times S^5$," JHEP 
\textbf{07} (2004) 004 [hep-th/0405102];
L. Freyhult, ``Bethe ansatz and fluctuations in SU(3) Yang-Mills 
operators," JHEP \textbf{06} (2004) 010 [hep-th/0405167];
J.A. Minahan, ``Higher loops beyond the SU(2) sector," 
JHEP \textbf{10} (2004) 053 [hep-th/0405243];
C. Kristjansen and T. Mansson, ``The circular, elliptic
three-spin string from the SU(3) spin chain," 
Phys. Lett. \textbf{B596} (2004) 265 [hep-th/0406176].
\bibitem{KL} A. Khan and A.L. Larsen, ``Spinning pulsating
solitons in $AdS_5 \times S^5$," Phys. Rev. \textbf{D69} (2004)
026001 [hep-th/0310019]; 
A.L. Larsen and A. Khan, ``Novel explicit 
multi spin string solitons in $AdS_5$," Nucl. Phys. \textbf{B686}
(2004) 75 [hep-th/0312184];
S. Ryang, ``Folded three-spin string solutions in
$AdS_5 \times S^5$," JHEP \textbf{04} (2004) 053 [hep-th/0403180];
H. Dimov and R.C. Rashkov, ``Generalized pulsating strings,"
JHEP \textbf{05} (2004) 068 [hep-th/0404012];
A. Khan and A.L. Larsen, ``Improved stability for pulsating
multi-spin string solitons," hep-th/0502063;
P. Bozhilov, ``String solutions in general backgrounds,"
hep-th/0503026.
\bibitem{AM} A. Mikhailov, ``Speeding strings," JHEP \textbf{12} (2004)
058 [hep-th/0311019];
``Slow evolution of nearly-degenerate extremal surfaces,"
hep-th/0402067; ``Supersymmetric null-surfaces," JHEP \textbf{09} 
(2004) 068 [hep-th/0404173]; 
``Notes on fast moving strings," hep-th/0409040;
``Plane wave limit of local conserved charges," hep-th/0502097.
\bibitem{MK} M. Kruczenski, ``Spin chains and string theory,"
Phys. Rev. Lett. \textbf{93} (2004) 161602 [hep-th/0311203]. 
\bibitem{KRT} M. Kruczenski, A.V. Ryzhov and A.A. Tseytlin, ``Large
spin limit of $AdS_5 \times S^5$ string theory and low energy expansion
of ferromagnetic spin chains," Nucl. Phys. \textbf{B692} (2004) 3
[hep-th/0403120];
M. Kruczenski and A.A. Tseytlin, ``Semiclassical
relativistic strings in $S^5$ and long coherent operators in 
$\mathcal{N}=4$ SYM theory," JHEP \textbf{09} (2004) 038
[hep-th/0406189].
\bibitem{DR} H. Dimov and R.C. Rashkov, ``A note on spin chain/string
duality," hep-th/0403121;
R. Hernandez and E. Lopez, ``The SU(3) spin chain sigma
model and string theory," JHEP \textbf{04} (2004) 052 [hep-th/0403139];
B. Stefanski, jr. and A.A. Tseytlin, ``Large spin limits
of AdS/CFT and generalized Landau-Lifshitz equations," 
JHEP \textbf{05} (2004) 042 [hep-th/0404133];
A.V. Ryzhov and A.A. Tseytlin, ``Towards the exact dilatation operator
of $\mathcal{N}=4$  super Yang-Mills theory," Nucl. Phys. 
\textbf{B698} (2004) 132 [hep-th/0404215];
K. Ideguchi, ``Semiclassical strings on $AdS_5 \times 
S^5/Z_M$ and operators in orbifold field theories," JHEP \textbf{09}
(2004) 008 [hep-th/0408014];
S. Bellucci, P.-Y. Casteill, J.F. Morales, C. Sochichiu,
``SL(2) spin chain and spinning strings on $AdS_5 \times S^5$,"
Nucl. Phys. \textbf{B707} (2005) 303 [hep-th/0409086];
S. Ryang ``Circular and folded multi-spin strings in spin chain
sigma models," JHEP \textbf{10} (2004) 059 [hep-th/0409217];
R. Hernandez and E. Lopez, ``Spin chain sigma models with fermions,"
JHEP \textbf{11} (2004) 079 [hep-th/0410022];
Y. Susaki, Y. Takayama and K. Yoshida, ``Open semiclassical strings
and long defect operators in AdS/dCFT correspondence," hep-th/0410139;
S. Bellucci, P.-Y. Casteill and J.F. Morales, ``Superstring sigma
models from spin chains: the SU(1,1$|$1) case," hep-th/0503159;
B. Stefanski, jr. and A.A. Tseytlin, ``Super spin chain coherent state
actions and $AdS_5 \times S^5$ superstring," hep-th/0503185.
\bibitem{AAT} A.A. Tseytlin, ``Semiclassical strings and 
AdS/CFT," in: Proceedings of NATO Advanced Study Institute and EC
Summer School on String Theory: ``From Gauge Interactions to Cosmology,"
Carges, France, 7-19 Jun 2004. hep-th/0409296.
\bibitem{CLM} C.G. Callan, H.K. Lee, T. McLoughlin, J.H. Schwarz, 
I. Swanson and X. Wu, ``Quantizing string theory in $AdS_5 \times S^5$:
Beyond the pp-wave," Nucl. Phys. \textbf{B673} (2003) 3 [hep-th/0307032];
C.G. Callan, T. McLoughlin and I. Swanson, ``Holography beyond the 
Penrose limit," Nucl. Phys. \textbf{B694} (2004) 115 [hep-th/0404007]; 
C.G. Callan, T. McLoughlin, and I. Swanson, ``Higher impurity AdS/CFT
correspondence in the near-BMN limit," Nucl. Phys. \textbf{B700} 
(2004) 271 [hep-th/0405153].
\bibitem{PR} A. Parnachev and A.V. Ryzhov, ``Strings in the near
plane wave background and AdS/CFT," JHEP \textbf{10} (2002) 066 
[hep-th/0208010].
\bibitem{BDS} N. Beisert, V. Dippel and M. Staudacher, ``A novel long
range spin chain and planar $\mathcal{N} =4$ super Yang-Mills,"
JHEP \textbf{07} (2004) 075 [hep-th/0405001].
\bibitem{AFS} G. Arutyunov, S. Frolov and M. Staudacher, ``Bethe
ansatz for quantum strings," JHEP \textbf{10} (2004)
016 [hep-th/0406256].
\bibitem{MS} M. Staudacher, ``The factorized S-matrix of CFT/AdS,"
hep-th/0412188.
\bibitem{NBT} N. Beisert, ``Spin chain for quantum strings," 
hep-th/0409054. 
\bibitem{AF} G. Arutyunov and S. Frolov, ``Integrable Hamiltonian
for classical strings on $AdS_5 \times S^5$," JHEP 
\textbf{02} (2005) 059 [hep-th/0411089].
\bibitem{IS} I. Swanson, ``On the integrability of string theory in
$AdS_5 \times S^5$," hep-th/0405172; ``Quantum string integrability
and AdS/CFT," Nucl. Phys. \textbf{B709} (2005) 443 [hep-th/0410282].
\bibitem{BPR} I. Bena, J. Polchinski and R. Roiban, ``Hidden symmetries
of the $AdS_5 \times S^5$ superstring," Phys. Rev. \textbf{D69} 
(2004) 046002 [hep-th/0305116];
N. Beisert, V.A. Kazakov, K. Sakai and K. Zarembo,
``The algebraic curve of classical superstrings on $AdS_5 \times S^5$,"
hep-th/0502226; ``Complete spectrum of long operators in
$\mathcal{N}=4$ SYM at one loop," hep-th/0503200;
L.A. Alday, G. Arutyunov and A.A. Tseytlin, ``On integrability of 
classical superstrings in $AdS_5 \times S^5$," hep-th/0502240;
J.A. Minahan, ``The SU(2) sector in AdS/CFT," hep-th/0503143.
\bibitem{LZ} M. L\"ubcke and K. Zarembo, ``Finite-size corrections to 
anomalous dimensions in $\mathcal{N} =4$ SYM theory," JHEP \textbf{05} 
(2004) 049 [hep-th/0405055]; 
S.A. Frolov, I.Y. Park and A.A. Tseytlin, ``On one-loop correction
to energy of spinning strings in $S^5$," Phys. Rev. \textbf{D71}
(2005) 026006 [hep-th/0408187];
I.Y. Park A. Tirziu and A.A. Tseytlin, ``Spinning strings in
$AdS_5 \times S^5$: One-loop correction to energy in SL(2) sector,"
hep-th/0501203;
L. Freyhult and C. Kristjansen, ``Finite size corrections to
three-spin string duals," hep-th/0502122;
N. Beisert, A.A. Tseytlin and K. Zarembo, ``Matching quantum strings
to quantum spins: one-loop vs. finite-size corrections," hep-th/0502173;
R. Hernandez, E. Lopez, A. Perianez and G. Sierra, ``Finite size
effects in ferromagnetic spin chains and quantum corrections to
classical strings," hep-th/0502188.
\bibitem{MKI} M. Kruczenski, ``Spiky strings and single trace 
operators in gauge theories," hep-th/0410226.

\end{thebibliography}
\end{document}